\documentclass[aps,prl,twocolumn,superscriptaddress,showpacs,amsmath,nofootinbib]{revtex4}
\usepackage{graphicx}
\usepackage{bm}

\def\lsim{\mathrel{\rlap{\lower4pt\hbox{\hskip1pt$\sim$}}
    \raise1pt\hbox{$<$}}}         
\def\gsim{\mathrel{\rlap{\lower4pt\hbox{\hskip1pt$\sim$}}
    \raise1pt\hbox{$>$}}}         

\begin{document}


\title{A Long, Cold, Early $r$-process? 
$\nu$-induced Nucleosynthesis in He Shells Revisited}



\author{Projjwal Banerjee}
\email{banerjee@physics.umn.edu}
\affiliation{School of Physics and Astronomy, University of Minnesota, Minneapolis, MN 55455}
\author{W. C. Haxton}
\email{haxton@berkeley.edu}
\affiliation{Department of Physics, University of California, and Lawrence Berkeley National Laboratory,
Berkeley, CA 94720}
\author{Yong-Zhong Qian}
\email{qian@physics.umn.edu}
\affiliation{School of Physics and Astronomy, University of Minnesota, Minneapolis, MN 55455}


\date{\today}

\begin{abstract}
We revisit a $\nu$-driven $r$-process mechanism
in the He shell of a core-collapse supernova, finding that it 
could succeed in early stars of metallicity 
$Z\lesssim 10^{-3}Z_\odot$, at relatively low
temperatures and neutron densities, producing
$A\sim 130$ and 195 abundance peaks over $\sim 10$--20~s.  The mechanism
is sensitive to the $\nu$ emission model and to $\nu$ oscillations.  
We discuss the implications of an $r$-process that could alter interpretations
of abundance data from metal-poor stars, and point out the need
for further calculations that include effects of the supernova shock.
\end{abstract}


\maketitle


While the basic features of the rapid-neutron-capture or $r$-process 
have been known for over 50 years 
\cite{BBFH}, the search for the specific astrophysical site 
has frustrated many researchers \cite{Cowan}.  The situation has continued despite
a growing set of observational constraints, including elemental
abundances from metal-poor (MP) stars \cite{Sneden}, that
appear to favor core-collapse supernovae (SNe) and to
disfavor some otherwise attractive sites, such as neutron star mergers (NSMs)
\cite{Mathews,Qian}. 

The surface compositions of old MP stars provide a fossil record of
nucleosynthesis and chemical enrichment in the early Galaxy. 
For ultra-metal-poor (UMP) stars, where
 [Fe/H]~$\equiv\log({\rm Fe/H})-\log({\rm Fe/H})_\odot\lesssim -3$, surface enrichments
should reflect contributions from just a few nearby nucleosynthetic events. 
The data show that the $r$-process operated
in the early Galaxy with a frequency consistent with SNe from short-lived 
massive progenitors. Many MP stars, including several UMP ones,
also exhibit a solar-like abundance pattern of heavy 
$r$-process elements ($r$-elements) for $A>130$ \cite{Sneden}.

The similarity between the MP-star and solar $r$-patterns
tempts one to conclude that there is a unique site for the $r$-process,
operating unchanged over the Galaxy's history (cf. \cite{WBG}). 
But is this the case?
Epstein, Colgate, and Haxton (ECH) \cite{ECH} suggested a possible 
$r$-site some years ago that would complicate such
an interpretation.  The ECH mechanism utilizes  
neutrons produced by neutral-current (NC) $\nu$
reactions in the He zones of certain low-metallicity SNe.
The proposed sequences are
$^4{\rm He}(\nu,\nu n)^3{\rm He}(n,p)^3{\rm H}(^3{\rm H},2n)^4{\rm He}$ and
$^4{\rm He}(\nu,\nu p)^3{\rm H}(^3{\rm H},2n)^4{\rm He}$.
For temperatures $\lesssim 3 \cdot 10^8$~K, the neutrons thus produced 
will not reassemble
into $^4$He by reactions involving light nuclei. Nor will they be captured 
by $^4$He as $^5$He is unbound. Instead, they will
be efficiently captured by seed nuclei, such as $^{56}$Fe, present in 
the birth material of the SN. The ECH neutron source is primary and
provides a roughly fixed number of neutrons.
For MP progenitors there are few Fe seeds and thus enough neutrons per seed
to produce heavy $r$-elements. As the metallicity of 
the SN increases, the neutron/seed ratio decreases, 
limiting the production of $r$-elements to low $A$ and eventually 
stopping the production altogether. That is, the ECH mechanism 
turns off with increasing metallicity.

The ECH mechanism was proposed as a candidate general $r$-process, and
thus was critiqued in Ref.~\cite{WHHH90} for being viable 
only in low-metallicity, compact SNe. Subsequent re-examination of
the mechanism focused on NC $\nu$ reactions only, either confirming
earlier results or finding no significant production of $A>80$ nuclei without
assuming ad hoc conditions in outer He zones \cite{NPB}.
In this Letter we show that the charged-current (CC) reaction
$^4$He$(\bar\nu_e,e^+n)^3$H can be an efficient neutron source for
a successful low-metallicity ECH mechanism using recently generated models
of MP massive stars \cite{WHW2002}.  Because other candidate $r$-sites,
such as NSMs, may turn on at higher metallicity, it is 
clearly important to explore any mechanism that might account for the $r$-elements
generated at earlier times.  Furthermore, as we have so far failed to identify
``the $r$-process," it would be a step forward to identify ``an $r$-process,'' even if 
the mechanism operated only for a limited time.

An $r$-process requires neutron densities  $n_n\gtrsim 10^{18}$ /cm$^3$, so that neutron capture
will be fast compared to $\beta$ decay, and a neutron/seed ratio $\gtrsim$ 80, so
that heavy $r$-elements can be produced from seeds like $^{56}$Fe. These requirements
lead us to examine the outer He shells of MP massive stars, where the low abundances
of nuclei like $^{12}$C, $^{14}$N, and $^{16}$O make iron-group nuclei an important neutron sink.
(The higher temperatures found in the inner He zone,  $\sim 3 \cdot 10^8$~K, lead to significant
$^{12}$C and $^{16}$O production by He burning, regardless of metallicity.  As we discuss later,
a modified ECH mechanism may operate in such an environment, with $\nu$-induced neutrons
``banked" in $^{13}$C and $^{17}$O, then liberated on shock wave passage.)

We use models u11--u75 of 11--$75\,M_\odot$ stars with an initial metallicity $Z=10^{-4}Z_\odot$
($Z$ being the total mass fraction of elements heavier than He) presented in
Ref.~\cite{WHW2002}. The outer He shells of these models
are at radii $r\sim 10^{10}$~cm, for which the gravitational collapse time is
\begin{equation}
\tau_{\rm coll}\sim\frac{1}{\alpha}\sqrt{\frac{r^3}{2GM}} 
\sim 102\left( \frac{0.6}{\alpha}\right)\left(\frac{M_\odot}{M}\right)^{1/2}r_{10}^{3/2}~\mathrm{s},
\end{equation}
where $\alpha \sim 0.6$ is the ratio of the infall velocity to the free-fall velocity,
$M\sim 2.4$--$33\,M_\odot$ is the mass enclosed within $r$, and $r_{10}$ is $r$ in units of $10^{10}$~cm. 
For such large $\tau_{\rm coll}$, we can assume that the radius, density, and temperature
of the He-shell material stay constant before the SN shock arrives.
We take the time of shock arrival to be approximately given by the Sedov solution \cite{WHHH90}
\begin{equation}
\tau_{\rm sh}\sim 21.8\left(\frac{M-M_{\rm NS}}{M_\odot}\right)^{1/2}\frac{r_{10}}{E_{50}^{1/2}}~\mathrm{s},
\label{eq:Sedov}
\end{equation}
where $M_{\rm NS}\sim 1.4\,M_\odot$ is the mass of the neutron star produced by the core
collapse and $E_{50}$ is the explosion energy in units of $10^{50}$~ergs. Following the passage 
of the shock, both the temperature and density of the material first increase rapidly and then 
decrease on timescales comparable to $\tau_{\rm sh}$. The peak temperature
(in units of $10^8$~K) of the shocked material is \cite{WHHH90}
\begin{equation}
T_{p,8} \sim 2.37E_{50}^{1/4}r_{10}^{-3/4}.
\end{equation}
For such low temperatures, photo-dissociation of heavy nuclei will not occur \cite{WHHH90}.  Other
effects of shock-wave passage are helpful to the $r$-process (see discussion below).

During the several seconds following core collapse, an intense flux 
of $\nu$s irradiates the He zone.  While the zone's radius, density, and temperature are unchanged, 
$\nu$ reactions must induce and maintain a free-neutron
density $n_n\gtrsim 10^{18}$/cm$^{3}$ to drive an $r$-process.  
We take the $\nu$ luminosity to be $L_\nu(t)=L_\nu(0)\exp(-t/\tau_\nu)$ for each of the six flavors,  
with $L_\nu(0)=1.67\cdot 10^{52}$~erg/s and $\tau_\nu=3$~s, so that the total energy
carried off by $\nu$s is $3\cdot 10^{53}$ ergs.
We use Fermi-Dirac $\nu$ spectra with zero chemical potential. 
We adopt nominal temperatures 
$T_{\nu_e}$, $T_{\bar\nu_e}$, and $T_{\nu_x}$
of 4, 5.33, and 8~MeV,
respectively, where $\nu_x$ stands for any heavy flavor, but explore the temperature dependence. 
Our nominal parameters are typical
of earlier SN models (e.g., \cite{WW94}).  The spectra at the He zone
will be affected by $\nu$ oscillations \cite{Lunardini}, as the $\nu$ mass splitting
$|\delta m_{13}^2| \sim 2.4\cdot 10^{-3}$~eV$^2$ produces a level crossing for a 20 MeV
$\nu$ at $\rho \sim 1.6 \cdot 10^3$ g/cm$^3$, a density characteristic of the carbon zone.  
The consequences for the $r$-process depend critically on the assumed $\nu$ mass hierarchy.

We evaluated the nucleosynthesis for models u11--u75 and for various $\nu$ oscillation
scenarios.  As an example of a successful $r$-process, we present detailed results for 
zone 597 of u11, assuming an inverted $\nu$ mass hierarchy (IH,
full $\bar\nu_e \leftrightarrow \bar\nu_x$ conversion).  Zone parameters are $r_{10}=1.10$, 
$M=2.43\,M_\odot$, $\rho=50.3$~g/cm$^{3}$, and $T_8=0.848$.  The zone
is nearly pure $^4$He: the initial mass fractions of $^{12}$C and $^{14}$N are
$X_{12} \sim1.39\cdot 10^{-5}$ and $X_{14} \sim1.35\cdot 10^{-6}$.  The total mass 
fraction of $A\ge16$ nuclei is $\sim 3.52\cdot 10^{-7}$ ($\sim 3.15\cdot 10^{-8}$ from $^{56}$Fe).
A big bang nucleosynthesis network \cite{BBN} was modified to 
follow the ECH mechanism, with NC and CC $\nu$ cross sections 
taken from Ref.~\cite{Gazit}, which agree well with those of Ref.~\cite{ECH}.
As the network stops at $^{16}$O, neutron capture on $A\ge16$ nuclei was approximated by a
constant loss rate corresponding to the initial abundances of such nuclei.  As discussed below,
the evolution of the neutron number fraction $Y_n$ is not significantly altered by neglecting
changes in the $A\ge16$ composition.

\begin{figure}
\includegraphics[width=8cm]{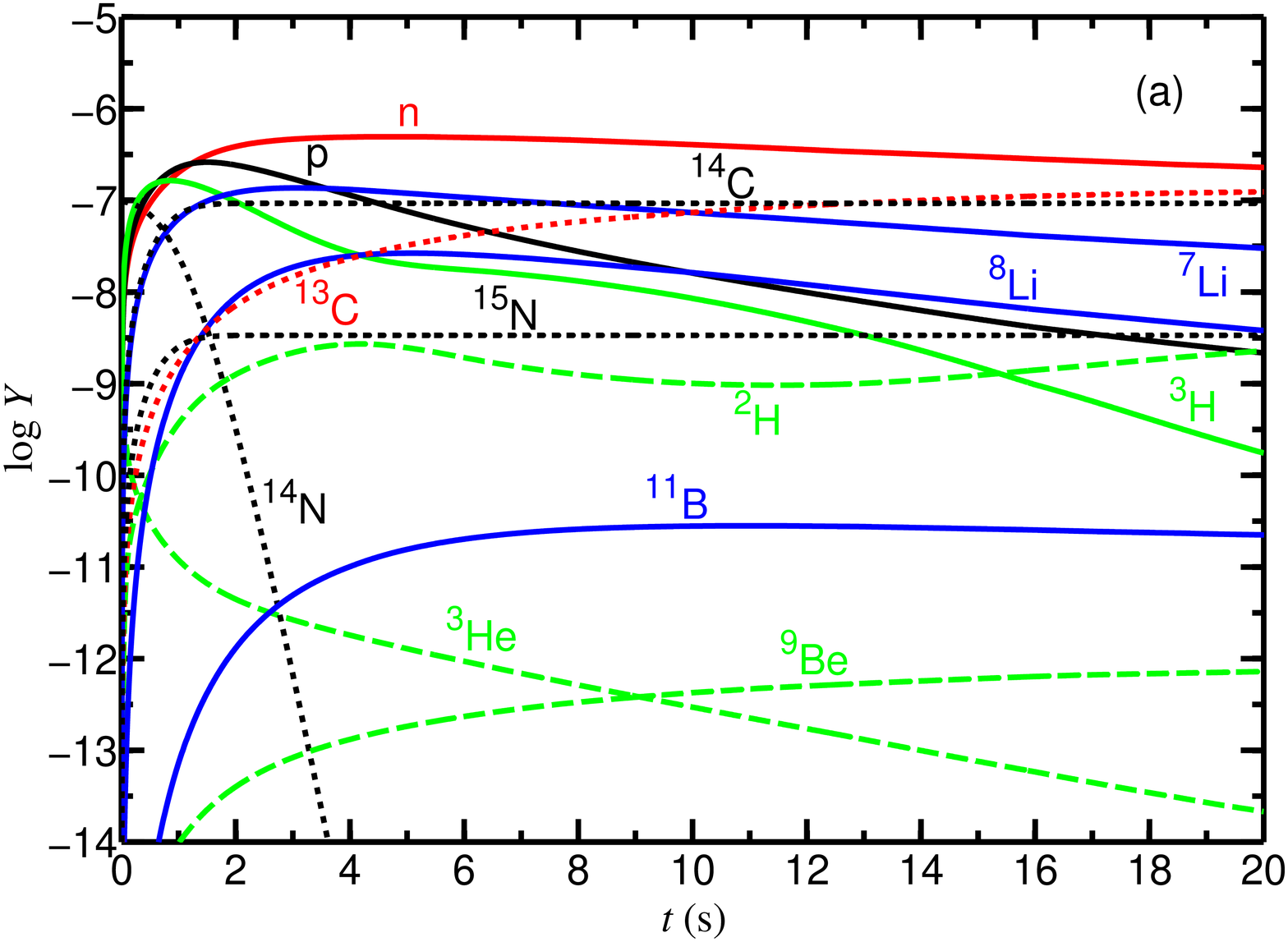}
\includegraphics[width=8cm]{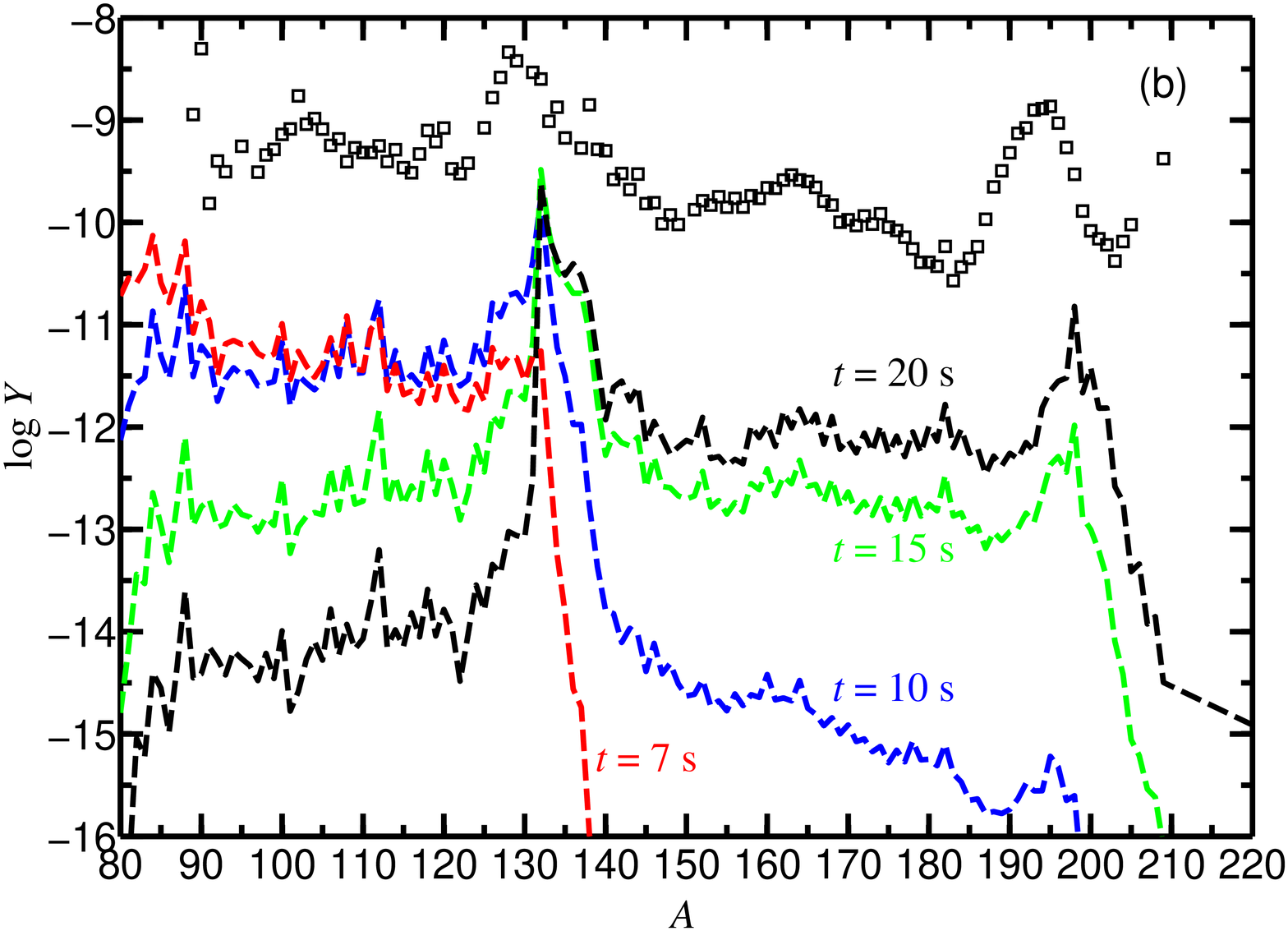}
\caption{$\nu$-induced nucleosynthesis in u11, zone 597:
(a) Number fractions $Y_i(t)$ of $A<16$ nuclei;
(b) $r$-process yields at $t=7$, 10, 15, and 20~s compared to solar $r$-pattern (squares).}
\label{fig:nur}
\end{figure}

Figure~\ref{fig:nur}a, the number-fraction evolution with time $t$, can be readily understood:
(1) The extremely efficient reaction $^3$He$(n,p)^3$H immediately consumes all neutrons produced
by the NC reaction $^4$He$(\nu,\nu n)^3$He.  Each NC reaction thus yields one proton and
one $^3$H.  (2) The neutron-producing reaction proposed by ECH, $^3$H$(^3$H,$2n)^4$He, is
inefficient. Instead, $^3$H is destroyed by abundant $^4$He via
$^3$H$(^4$He,$\gamma)^7$Li.  Neutron restoration by $^7$Li$(^3$H,$2n)2^4$He is ineffective
for the conditions of Figure~\ref{fig:nur}a. (3) Neutron production is dominated by the CC reaction 
$^4$He$(\bar\nu_e,e^+n)^3$H. (4) The principal neutron sinks are $^7$Li, $^{12}$C, and $A\ge16$
nuclei.  (5) Protons are not a significant neutron sink 
as $p(n,\gamma)^2$H is immediately followed by $^2$H$(^3$H,$n)^4$He. (6) Due to its
small initial abundance, neutron capture by $^{14}$N is also negligible.

The rate of the CC $\bar\nu_e$ reaction per $^4$He nucleus is
\begin{equation}
\lambda_{\bar\nu_e\alpha}^{\rm CC}(t)=\frac{2.28\times10^{-7}}{r_{10}^2\exp(t/\tau_\nu)}
\left(\frac{T_{\bar\nu_e}}{6\ {\rm MeV}}\right)^k\ {\rm s}^{-1},
\label{eq:cc}
\end{equation}
where $k \sim 6.26$ and  $\sim 5.17$ for $T_{\bar\nu_e}=4$--6 and 6--8~MeV, respectively.
Based on the above discussion, $Y_n$ in Figure~\ref{fig:nur}a can be estimated from
\begin{equation}
\dot Y_n=\lambda_{\bar\nu_e\alpha}^{\rm CC}(0)Y_\alpha\exp(-t/\tau_\nu)
-\lambda_{n,\gamma}Y_n(t),
\label{eq:dyndt}
\end{equation}
where $\lambda_{\bar\nu_e\alpha}^{\rm CC}(0)=8.35 \cdot10^{-7}$/s
for $T_{\bar\nu_e}=8$~MeV (IH), $Y_\alpha \sim 1/4$ is the number fraction
of $^4$He, and $\lambda_{n,\gamma} \sim 8.12\times 10^{-2}$/s is the net rate 
of neutron capture on $^7$Li (46.2\%), $^{12}$C (21.9\%), and $A\ge16$ nuclei (31.9\%). We find,
in good agreement with Figure~\ref{fig:nur}a,
\begin{equation}
Y_n(t)=\frac{\lambda_{\bar\nu_e\alpha}^{\rm CC}(0)Y_\alpha\tau_\nu}
{1-\lambda_{n,\gamma}\tau_\nu}[\exp(-\lambda_{n,\gamma}t)-\exp(-t/\tau_\nu)].
\label{eq:ynt}
\end{equation}

\begin{figure}
\includegraphics[width=8cm]{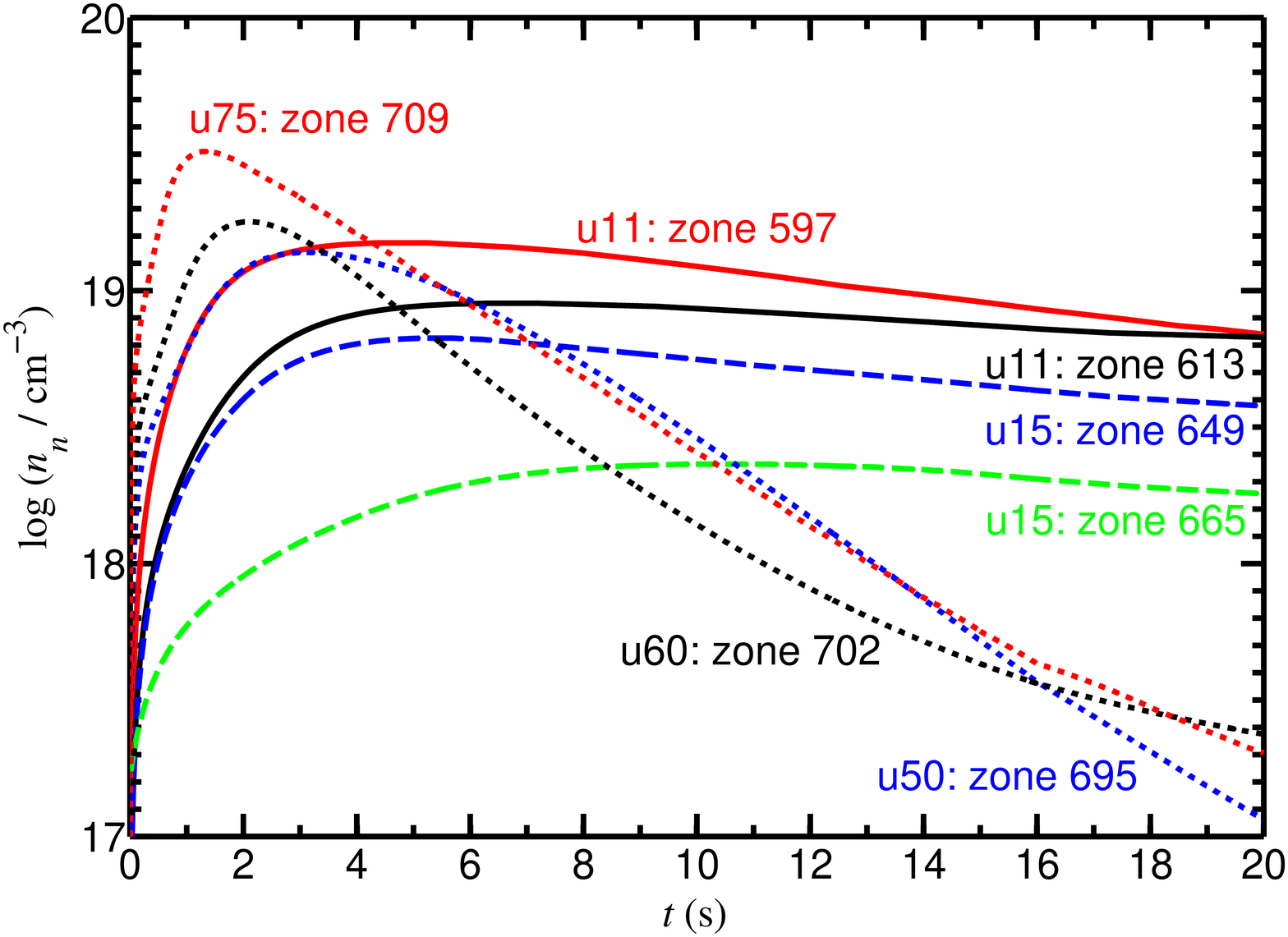}
\caption{Neutron number density $n_n(t)$ evolution for selected outer He zones in 
models u11, u15, u50, u60, and u75.}
\label{fig:nn}
\end{figure}

The neutron number density  in zone 597 of u11,
$n_n=Y_n\rho N_A\sim 10^{19}$/cm$^{3}$ where $N_A$ is Avogadro's number,
is sufficient to drive an $r$-process 
(see Figure~\ref{fig:nn}).
The most effective seed is $^{56}$Fe as it is above the $N=28$ closed neutron shell.
The typical mass number of $r$-elements produced at time $t$ is roughly
$A\sim 56+N_{\rm cap}(t)$, where
$N_{\rm cap}(t)=\int_0^tn_n(t')\langle v\sigma_{n,\gamma}({\rm Fe})\rangle dt'$ and where
$\langle v\sigma_{n,\gamma}({\rm Fe})\rangle$ is
the rate coefficient for neutron capture on $^{56}$Fe. 
For zone 597 we find 
$N_{\rm cap}(t)=88$ (226) for $t=7$ (20)~s, which correspond to
the shock arrival times for $E_{50}\sim 12$ (1). We conclude, for weak explosions,
that the $r$-process could run to completion in the pre-shock phase.

We followed the nuclear flow from $^{56}$Fe with a large network Torch \cite{torch} 
that includes all of the relevant neutron capture, photo-disintegration, and $\beta$-decay
reactions. The yields at $t=7$, 10, 15, and 20~s are shown in Figure~\ref{fig:nur}b
along with the scaled solar $r$-pattern. The $r$-process is cold: 
photo-disintegration is unimportant for He zone temperatures.  It is also much slower
than usually envisioned.
At $t=7$~s, the $r$-process flow barely reaches the $A\sim 130$ peak.
Significant production of nuclei with $A>130$ occurs only
for $t>10$~s, and formation of a significant peak at $A\sim 195$ requires $t\sim 20$~s.
These times are readily understood.  The peaks at $A\sim 130$ and 195
correspond to parent nuclei $\sim$
$^{130}$Cd and $\sim$ $^{195}$Tm with closed neutron shells of $N=82$ and 126. 
With $^{56}$Fe as the seed, 74 neutron-capture and 22 $\beta$-decay
reactions are required to reach $^{130}$Cd while 139 neutron-capture and 43 $\beta$-decay
reactions are required to reach $^{195}$Tm.  In the absence of photo-disintegration, the
$r$-path is governed by $(n,\gamma)$-$\beta$ equilibrium and the rates for neutron capture and
$\beta$ decay will be comparable. For $\langle v\sigma_{n,\gamma}({\rm Fe})\rangle\sim
10^{-18}$~cm$^3$/s and $n_n\sim 10^{19}$/cm$^{3}$, the neutron-capture rate on 
$^{56}$Fe is $\sim 10$/s. As this rate is typical along the $r$-path,
$^{130}$Cd and $^{195}$Tm will be reached in $\sim 10$ and 18~s.

We examined other u11 zones and other progenitors. For the IH case with
$T_{\bar{\nu}_e} \sim 8$ MeV,
neutron densities of $\sim 10^{18}$--$10^{19}$/cm$^{3}$
are produced in many zones of models u11--u16 and u49--u75.   Conditions in u11--u16 are 
similar to those of zone 597 of u11, but the u49--u75 zones are hotter and denser,
$T_8\sim 2$--3 and $\rho\sim 200$--600~g/cm$^{3}$.  Figure~\ref{fig:nn} shows
$n_n(t)$ for selected zones of u11, u15, u50, u60, and u75. 
A much higher rate of neutron capture in u50, u60, and u75 leads to more rapid 
decline of $n_n(t)$.  Substantial $r$-yields are expected in the outer He zones
of 11--16 and 49--$75\,M_\odot$ stars at $Z \sim 10^{-4}Z_\odot$.  An $r$-process is not
expected for stars between 17 and $48\,M_\odot$ because the outer He zone
has too much hydrogen, a neutron poison.

The total yield of heavy $r$-elements from each SN is $\Delta M_r\sim 10^{-8}\,M_\odot$,
comparable to $\sim 4\cdot10^{-8}\,M_\odot$ in the Sun. Abundances of heavy $r$-elements
in MP stars with [Fe/H]~$<-2.5$ are $\sim 3\cdot10^{-4}$--$10^{-1}$ times those in the Sun
\cite{Sneden}. At least some $r$-enrichments in this range could be produced by an SN in
the early interstellar medium, but this process then turns off as progenitor metallicity increases.
Both $n_n(t)$ and the $A>56$ yields decrease significantly 
with increasing progenitor $Z$.  In the scenarios studied here,  $r$-process conditions
are not found beyond $Z \sim 10^{-3}Z_\odot$.   Yet net neutron production by $\nu$s 
is insensitive to metallicity, depending only on SN energy, $\bar{\nu}_e$
temperature, and shell radius,  so neutron capture continues on stable seeds like
$^{56}$Fe, modestly increasing the $A>56$ yields.  The net
mass of heavy nuclei continues to be  incremented by $\sim 10^{-8}\,M_\odot$.
The associated Galactic chemical evolution \cite{BQH2} should be studied to determine how
the $\nu$-driven mechanism might merge into other $r$-processes, such as NSMs, 
that may only be viable for [Fe/H]~$\gtrsim-2.5$ \cite{Qian}.  

We have used two separate networks to estimate $n_n(t)$ and the corresponding 
$r$-yields. In estimating $n_n(t)$, we adopt a constant neutron capture rate
for $A \ge 16$ nuclei.  This approximation should be valid because the important neutron sinks
$^7$Li and $^{12}$C are included, and because the
calculations confirm that the total number of neutrons captured per $^{56}$Fe nucleus 
is $\ll Y_n$. Nevertheless, future studies should use a complete network for both neutron 
capture and $\nu$ interactions. 

The effects of shock passage through the He shell have not been included, though we argued
that $r$-nuclei will survive the associated heating.  Other
consequences may be beneficial, extending the range for
interesting nucleosynthesis.   The density of shocked material jumps to $\sim 7$ times
the pre-shock value and then decreases slowly on timescales $\sim\tau_{\rm sh}$. 
So while larger explosion energies, $E_{50}\sim 12$, might appear to limit the duration
of the $r$-process to $\tau_{\rm sh}\sim 7$~s, in fact there may be a post-shock phase where
densities higher than those of  Fig.~\ref{fig:nn} aid the nucleosynthesis.  
Another potentially beneficial effect of the shock may come from neutrons
released by $^{13}$C$(^4$He,$n)^{16}$O and
$^{17}$O$(^4$He,$n)^{20}$Ne:  $^{12}$C and $^{16}$O are the principal neutron sinks
in the inner He shell.  If shock heating to $\gtrsim 5 \cdot 10^8$K could liberate these
neutrons without increasing the abundance of seeds, one might exploit both the more favorable
$1/r^2$ of the inner He zone and NC $\nu$ channels in neutron production (which in the
outer He zone lead to $^7$Li).  One source of uncertainty
comes from the $^{12}$C and $^{16}$O $(n,\gamma)$ cross sections,
which differ by factors of $\sim$ 3 and 45 (10 and 160) at $T_8 \sim 0.85$ 
(3) between Evaluated Nuclear Data File and Japanese Evaluated Nuclear Data Library
\cite{Prity}.  The differences reflect the energy range over which
s-wave capture is assumed to dominate.  Pending resolution of this discrepancy, 
parametric studies will be needed \cite{BQH2}.

The CC $\bar\nu_e$ reaction on $^4$He plays a crucial role in the $\nu$-induced
$r$-process presented here. The rate of this reaction is quite sensitive to the
$\bar\nu_e$ spectrum [see Eq.~(\ref{eq:cc})] and thus 
to both $\nu$ emission parameters and flavor oscillations.
For our adopted $\nu$ emission parameters,
only nuclei with $A\sim 70$--80 can be produced in the outer He zone
without oscillations, while no interesting nucleosynthesis occurs for the normal $\nu$ mass hierarchy
(strong $\nu_e \leftrightarrow \nu_x$ conversion). If we
lower $T_{\nu_x}$ from 8 to 6~MeV at emission, only nuclei with $A\sim 70$--80 can be 
produced even with full $\bar\nu_e \leftrightarrow \bar\nu_x$ conversion (IH). 
Recent SN simulations for 8.8--$18\,M_\odot$ progenitors yielded significantly softer 
$\nu$ spectra at emission than adopted above \cite{Hudepohl}.  In contrast,
spectra similar to ours were obtained for 
$\sim 40$--$50\,M_\odot$ progenitors associated with black-hole formation \cite{Sumi}.
Recent progress in SN modeling and in the nuclear microphysics governing $\nu$ opacity
is impressive and should encourage further efforts needed to determine $\nu$ temperatures
with small error bars.

In conclusion, we have explored one scenario for a cold $r$-process --- the $\nu$-driven He-shell
mechanism --- as a counterpoint to
more conventional high-temperature SN $r$-process mechanisms that typically run into 
problems of seed overgrowth.  
The $\nu$-induced mechanism is intriguing because it can be evaluated
quantitatively in realistic progenitors, and because it is remarkably sensitive to new $\nu$
physics.  We believe this cold, early mechanism merits investigation in other
astrophysical settings, including the inner He zone discussed above and 
the late stages of $\nu$-driven winds.   The mechanism could be part of a multiple-$r$-process
explanation of Galactic chemistry.
   
\begin{acknowledgments}
We thank Alexander Heger for discussions of massive stars
and Frank Timmes and Rob Hoffman for help with the Torch network.
This work was supported in part by the US DOE under DE-FG02-87ER40328 at UMN
and DE-SC00046548 at Berkeley.
\end{acknowledgments}

\end{document}